\documentclass[aps,prl,reprint,amsmath]{revtex4-1} 
\usepackage{times}
\usepackage{hyperref} 
\usepackage{graphicx}

\begin{document}
	
\title{Atomic White-Out: Enabling Atomic Circuitry Through Mechanically Induced Bonding of Single Hydrogen Atoms to a Silicon Surface}

\author{Taleana Huff$^{1,\ast,\dagger,3}$, Hatem Labidi$^{1,2,\ast}$, Mohammad Rashidi$^{1}$, Mohammad Koleini$^{1,2}$, Roshan Achal$^{1,3}$, Mark Salomons$^{2,3}$ and Robert A. Wolkow$^{1,2,\dagger,3}$}

\affiliation{$^1$Department of Physics, University of Alberta, Edmonton, Alberta, T6G 2J1, Canada}
\affiliation{$^2$National Institute for Nanotechnology, National Research Council of Canada, Edmonton, Alberta, T6G 2M9, Canada}
\affiliation{$^3$Quantum Silicon, Inc., Edmonton, Alberta, T6G 2M9, Canada}
\affiliation{$^{\ast}$These authors contributed equally to this work}
\affiliation{$\dagger$ email: taleana@ualberta.ca; rwolkow@ualberta.ca}

\begin{abstract}

We report the mechanically induced formation of a silicon-hydrogen covalent bond and its application in engineering nanoelectronic devices. We show that using the tip of a non-contact atomic force microscope (NC-AFM), a single hydrogen atom could be vertically manipulated. When applying a localized electronic excitation, a single hydrogen atom is desorbed from the hydrogen passivated surface and can be transferred to the tip apex as evidenced from a unique signature in frequency shift curves. In the absence of tunnel electrons and electric field in the scanning probe microscope junction at 0 V,  the hydrogen atom at the tip apex is brought very close to a silicon dangling bond, inducing the mechanical formation of a silicon-hydrogen covalent bond and the passivation of the dangling bond. The functionalized tip was used to characterize silicon dangling bonds on the hydrogen-silicon surface, was shown to enhance the scanning tunneling microscope (STM) contrast, and allowed NC-AFM imaging with atomic and chemical bond contrasts. Through examples, we show the importance of this atomic scale mechanical manipulation technique in the engineering of the emerging technology of on-surface dangling bond based nanoelectronic devices.

\end{abstract}

\maketitle

\section{Introduction}

Due to the continuous improvement of scanning probe microscopy techniques, the long thought inaccessible goal of inducing and visualizing chemical reactions at the atomic scale is now routinely achievable by many groups around the world.
In the framework of so-called mechanochemistry,\cite{Beyer.2005} mechanical force induced reactions have been studied using NC-AFM . \cite{Custance.2009} Recent works reported force induced atomic-scale switching,\cite{Sweetman.2011} quantitative force measurements to induce the diffusion of single atoms \cite{Ternes.2008} and molecules, \cite{Langewisch.2013} as well as studying molecular conformers \cite{Jarvis.2015} and tautomerization. \cite{Ladenthin.2016} A few earlier studies also showed examples of mechanically induced vertical manipulation of single atoms. \cite{Dujardin.1998,Oyabu.2003} However, direct observation of mechanically induced covalent bonding of two different atoms using NC-AFM remain scarce.\cite{Sugimoto.2005} 

Recently, the silicon dangling bond (DB) on the technologically relevant H-Si(100) surface was established as a very promising building block for beyond CMOS technology.\cite{Haider.2009,Wolkow.2014} A DB corresponds to a desorbed single hydrogen atom from the otherwise passivated silicon surface. It is approximately an $sp^3$ hybrid orbital that can be occupied by 2, 1, or 0 electrons resulting, respectively, in a negative, neutral, or positively charged DB. Thus, a DB behaves essentially as a single atom quantum dot, with charge state transitions reported in STM experiments.\cite{Taucer.2014,Labidi.2015}
DBs can be found natively on the surface as a result of imperfections during the hydrogen termination procedure or artificially created using the STM tip. Different works have shown that controlled atom-by-atom lithography, i.e. hydrogen desorption, on the H-Si surface allows creation of DB based circuits for next generation ultimately-miniaturized low power nanoelectronic devices. \cite{Haider.2009,Livadaru.2010,Schofield.2013,Wolkow.2014,Kolmer.2015} 

Although STM tip induced desorption of hydrogen from the H-Si(100) surface was extensively studied,\cite{Shen.1995,Foley.1998,Soukiassian.2003,Walsh.2009,Schmucker.2012,Schofield.2013,Moller.2017} the reverse manipulation of selective adsorption of a single hydrogen atom to passivate a silicon DB has not, to our knowledge, been reported so far. In this context, AFM can bring more insights by allowing identification of different tip dynamics \cite{Gross.2009,Sharp.2012} and probing chemical reactivity at the atomic scale. \cite{Lantz.2001,Sugimoto.2007}

Here, we report the first controlled vertical manipulation of a single H atom using the tip of an AFM sensor and its application in characterizing and engineering silicon DB-based structures of relevance to nanoelectronic devices. We show that following a localized tip induced excitation on the Si-H surface, a single hydrogen atom is desorbed and could be either deposited on the surface with stable imaging in STM and AFM, or transfered to the tip apex. The single H atom functionalized tip was identified through a unique signature in frequency shift vs. displacement curves (i.e. $\Delta f (z)$) and a characteristic enhancement of STM images in filled and empty states.
By bringing the H-functionalized tip apex very close to a DB in the absence of bias and current, a covalent bond between the single hydrogen and silicon atoms is formed. Subsequent changes in the STM images and $\Delta f (z)$ curves confirmed that this mechanically induced reaction results in the passivation of the DB with the hydrogen from the tip apex. 

It has become clear that CO functionalized tips are effective for characterization of adsorbed molecules on metal surfaces.\cite{Gross.2009,Mohn.2011} It is clear also that accessible and effective tips are required for other systems of study. Preparing and identifying such a tip is described in this work. Moreover, the H functionalized tip is shown to allow characterization and also induce changes in DB-based structures on the H-Si(100) surface through selective mechanically induced hydrogen passivation, or "capping". 
 
\section{Results and discussions}
\begin{figure}[htbp]
	\centering
	\includegraphics[]{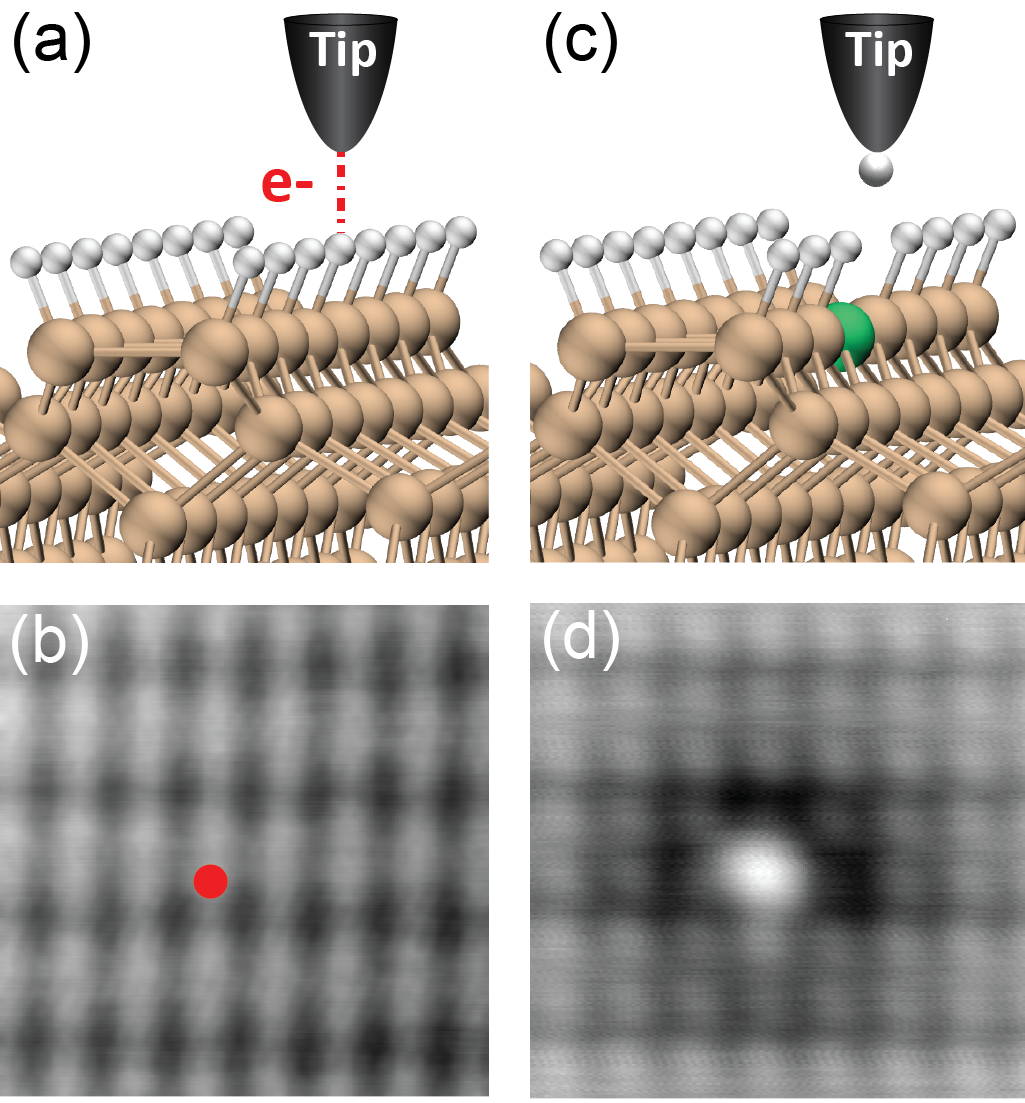}
	\caption{\textbf{An illustration of the tip induced manipulation that can result in tip functionalization with a single hydrogen atom.} (a) Ball and stick model of the H-Si(100)-2x1 surface. (b) Typical defect-free empty states STM image using a non-functionalized tip and showing the dimer structure of the surface. The red dot indicates the position of the STM tip when the electronic excitation sketched in (a) is applied. (c) Ball and stick model of a silicon dangling bond in green and a H-functionalized tip resulting from the tip-induced desorption. (d) Typical STM image of a DB acquired with a H-functionalized tip showing a characteristic STM contrast enhancement. Both STM images were acquired in constant current mode with a set point of 50~pA at +1.3~V.}
	\label{Fig1}
\end{figure}

\noindent \textbf{Tip functionalization with a single hydrogen atom.}
In the Si(100)-2$\times$1 reconstruction, silicon atoms at the surface are organized in dimers. When the surface is passivated with hydrogen in the monohydride reconstruction, each silicon atom at the surface is covalently bonded with a single hydrogen atom as represented in Figure \ref{Fig1}-a.
Figure \ref{Fig1}-b shows a typical defect-free empty states STM image acquired using a non functionalized tip (see methods and ref \cite{Labidi.2017} for details on in-situ tip preparation).

Figure \ref{Fig1}-c shows a 3D ball and stick model of a silicon dangling bond (represented in green) on the H-Si(100) surface. 
To create a single DB, the STM tip is positioned on top of a hydrogen atom (red dot in Figure \ref{Fig1}-b), then the feedback loop is switched off, and a voltage pulse of about 2.3~V is applied for a few milliseconds. As illustrated in Figure \ref{Fig1}-c, this results in the selective desorption of the hydrogen atom under the tip apex which is often transfered to the tip. 
Figure \ref{Fig1}-d shows a typical STM image of the created single DB. In accordance with earlier studies in the literature, the DB in empty states appears as a bright protrusion surrounded by a characteristic dark halo. \cite{Haider.2009,Labidi.2015}

\begin{figure}
	\centering
	\includegraphics[]{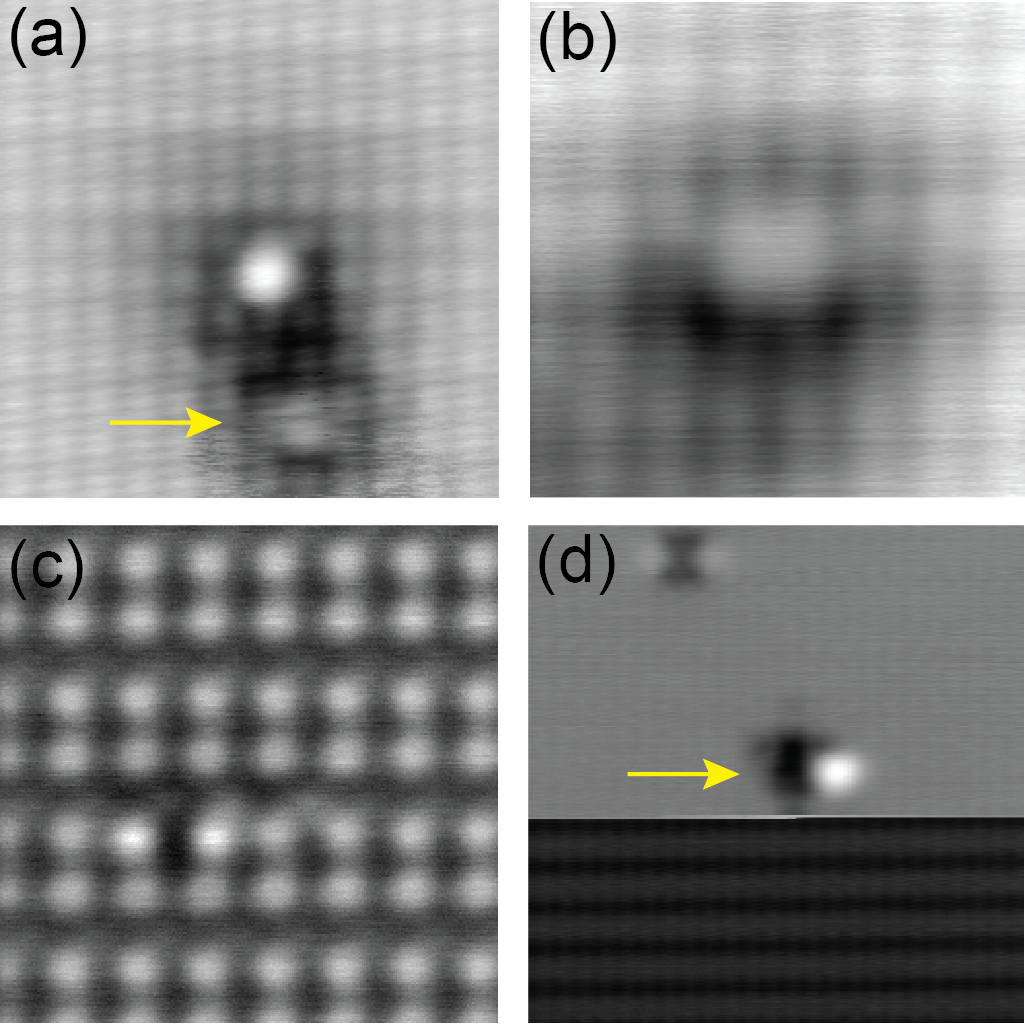}
	\caption{\textbf{Imaging a single hydrogen atom physisorbed on the H-Si(100) surface.}
		(a) A (5$\times$5)$nm^2$ STM image at +1.3~V of a DB where the desorbed atomic hydrogen was not picked up, instead adsorbing at the location indicated by an arrow. (b) (3$\times$3)$nm^2$ STM image of an atomic hydrogen adsorbed on the surface and (c) corresponding AFM frequency shift map at 0~V and a relative tip elevation of z= -3.8~\AA{}. (d) An atomic hydrogen on the surface is picked up by a slow downward STM scan at V=+1.6V. All STM images are constant current at 50~pA.}
	\label{Fig2}
\end{figure}

While the tip-induced desorption of hydrogen from the H-Si(100) surface has been studied by researchers for over two decades (see ref \cite{Labidi.2015} for details), the location and mode of attachment of the H atom after desorption has remained unknown. Following procedures described here and tracking desorption events for several different tips, we found that DB creation through a voltage pulse resulted in the desorbed H atom being transfered to the tip apex roughly 50\% of the times, i.e. forming a H-functionalized tip. In 30\% of cases, the desorbed H atom is found on the H-Si surface close to the just created DB, as shown in Figure \ref{Fig2}-a. In the remaining 20\% of cases, the tip apex does not change and a hydrogen atom could not be seen in the vicinity of the newly created DB, suggesting it was possibly adsorbed on the tip away from the apex atom, deposited on the surface farther from the DB, or ejected to the vacuum. 

Figure \ref{Fig2}-a shows an example of a single hydrogen atom found to be deposited on the H-Si surface immediately after tip induced creation of a DB. Such an object was confirmed to be a single hydrogen atom by dragging it with an elevated positive bias to passivate the created nearby DB (Supplementary materials, Figure S1). Interestingly, the hydrogen atom appears in empty states STM images as a slightly bright protrusion surrounded by a dark halo as shown in Figure \ref{Fig2}-b. This suggests a possible charging effect that induces a localized band bending similar to a single DB. \cite{Schofield.2013,Labidi.2015} In the corresponding frequency shift map (Figure \ref{Fig2}-c), the physisorbed hydrogen atom appears to induce a lattice distortion of two adjacent dimer pairs. When imaged at relatively high positive voltage (+1.6V in the example of Figure \ref{Fig2}-f), the hydrogen atom was picked up by the tip apex as evidenced from the change in STM contrast midway through the scan.

In the examples of Figure \ref{Fig1} and \ref{Fig2}, the enhanced STM contrast after creating a DB is a first strong indication of tip functionalization with the desorbed single H atom. The contrast changes from resolving dimers (Figure \ref{Fig1}-b) to resolving single atoms (Figure \ref{Fig1}-d), respectively, before and after the hydrogen desorption from the surface. This is similar to what is well known for the CO molecule, where once it is picked up by the tip apex following a voltage pulse it enhances the STM and AFM contrast.\cite{Bartels.1998,Gross.2009,Mohn.2013} In the following section, we provide further evidence of a single hydrogen atom functionalized tip using NC-AFM, which allows identification of different tip dynamics through studying force curves.\cite{Sharp.2012,Yurtsever.2013,Labidi.2017} 

\begin{figure*}
	\centering
	\includegraphics[width=\textwidth]{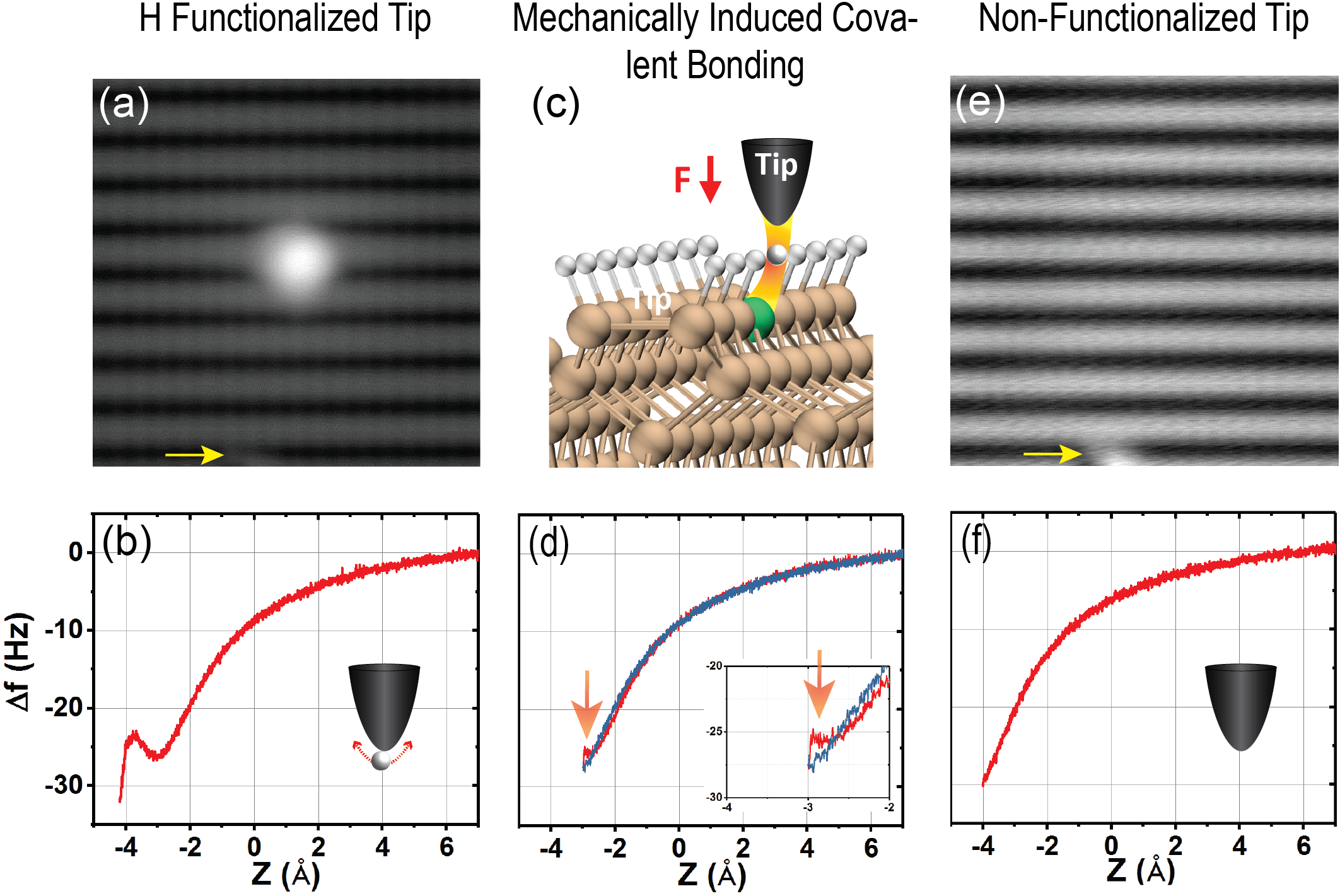}
	\caption{\textbf{Procedure to mechanically induce a hydrogen-silicon covalent bond.}
		(a) A typical filled states STM image of a silicon dangling bond on the H-Si(100)-2x1 surface using a single hydrogen atom functionalized tip. The yellow arrow indicates a defect taken as a reference.  (b) $\Delta$f(z) curve using H-functionalized tip on a surface hydrogen atom. (c) Ball and stick model and (d) $\Delta$f(z) curve on a single DB during the mechanically induced Si-H covalent bond capping event. The orange arrow indicates a hysteresis (zoom in inset) characteristic of the change that occurs due to the formation of the covalent bond between the H atom at the tip apex and the silicon dangling bond. (e) STM image and (f) $\Delta$f(z) curve on the H-Si surface subsequent to the mechanically induced reaction in (d).}
	\label{Fig3}
\end{figure*}

\noindent \textbf{Mechanically induced covalent bonding of single hydrogen and silicon atoms.}
Figure \ref{Fig3}-a shows a filled states STM image of the H-Si surface with a silicon DB created using the procedure described in the previous section. Similar to the case of empty states, we notice an enhanced STM contrast. In fact, typical filled states STM images of the H-Si surface usually show only dimer rows,\cite{Labidi.2015} but in Figure \ref{Fig3}-a the dimers of dimer rows are clearly resolved.

Following the creation of the single DB in Figure \ref{Fig3}-a, the scanner was switched to AFM scanning mode. Figure \ref{Fig3}-b shows a frequency shift vs. displacement curve recorded using a hydrogen functionalized tip on top of a hydrogen atom on the surface. The minima at around -3~\AA{} is always seen when a functionalized tip is prepared following the procedure previously described. Such features were reported by other works in the case of functionalized tips, and are ascribed to the relaxation of the functionalizing atom at the tip apex. \cite{Sugimoto.2008,Mohn.2011,Ladenthin.2016}

When recorded on a DB using the same functionalized tip, $\Delta$f(z) curves exhibit a hysteresis between the forward and backward sweep when the tip is brought very close to the DB as shown in Figure \ref{Fig3}-c and d, which indicates a change in the AFM junction.\cite{Sugimoto.2005,Sugimoto.2008,Berger.2015} When acquiring a subsequent STM image, we notice that the DB was capped with a hydrogen atom. The defect indicated by the yellow arrow is used as a marker showing that Figure \ref{Fig2}-e is exactly the same area as \ref{Fig2}-a. Additionally, $\Delta$f(z) curves recorded on top of a hydrogen atom of the surface as shown in Figure \ref{Fig3}-f no longer exhibit the minima characteristic of the hydrogen functionalized tip. This unambiguously proves that the tip that yields the minima in force curves was indeed functionalized with a single hydrogen atom.

Throughout all our experimental data, we saw that a tip that produces enhanced STM also systematically produces the characteristic force curves with the shallow minima. Therefore, change in the STM contrast, such as presented in Figures \ref{Fig1}-d, \ref{Fig2}-a, and \ref{Fig3}-(a) indicates successful functionalization of the tip apex with a single hydrogen atom. This is important for technological applications related to altering DB engineered structures through capping, as changes in STM contrast to detect H-functionalized tips is a much faster indicator than the time consuming acquisition of $\Delta$f(z) curves. In fact, regular systematic, non-tip-damaging, and reliable capping was produced using STM contrast as an indicator alone. 

All $\Delta$f(z) curves were recorded at 0~V in the complete absence of tunnel current, and the hydrogen capping of the DB only occurs when the tip is brought to a close enough interaction distance. Therefore, the silicon-hydrogen covalent bonding is mechanically induced. We note here that mechanically induced desorption was also observed, but often resulted in tip structure changes or multiple hydrogens desorbed, unlike the gentle and precise tip induced desorption. We note here, the initial tip apex structure before picking up a hydrogen atom on the tip apex is never exactly the same. So, the H-tip bond is not necessarily the same in all H-functionalized tips, similarly to the case of CO tips. This is likely the reason we observe variation on the tip elevation to induce capping. Other factors such as the sensor oscillation amplitude or the $\Delta$f(z) acquisition parameters might also play a role.

So far, reliable functionalization of tips mainly with CO molecules allowed different groups to achieve submolecular and bond contrast imaging of different molecular systems.\cite{Gross.2009,Mohn.2011} In the following, we show that in addition to high resolution AFM imaging, H-functionalized tips can be implemented in atom-by-atom lithography to create and modify silicon DB based nanoelectronic elements.\\

\noindent \textbf{Characterizing silicon dangling bonds on the Si-H surface with a H-functionalized tip.}
Although DBs on the H-Si surface have been extensively studied using STM, NC-AFM works remain almost nonexistent. 
AFM can provide an important complementary view to STM works as it allows characterizing the chemical reactivity of DBs. Moreover, unlike STM, AFM allows probing the electronic properties of DBs and DB structures in the band-gap with minimized perturbation from the tip, e.g. minimal tip induced band bending and electron/hole injection.\cite{Livadaru.2010,Labidi.2015} 

\begin{figure}
	\centering
	\includegraphics[]{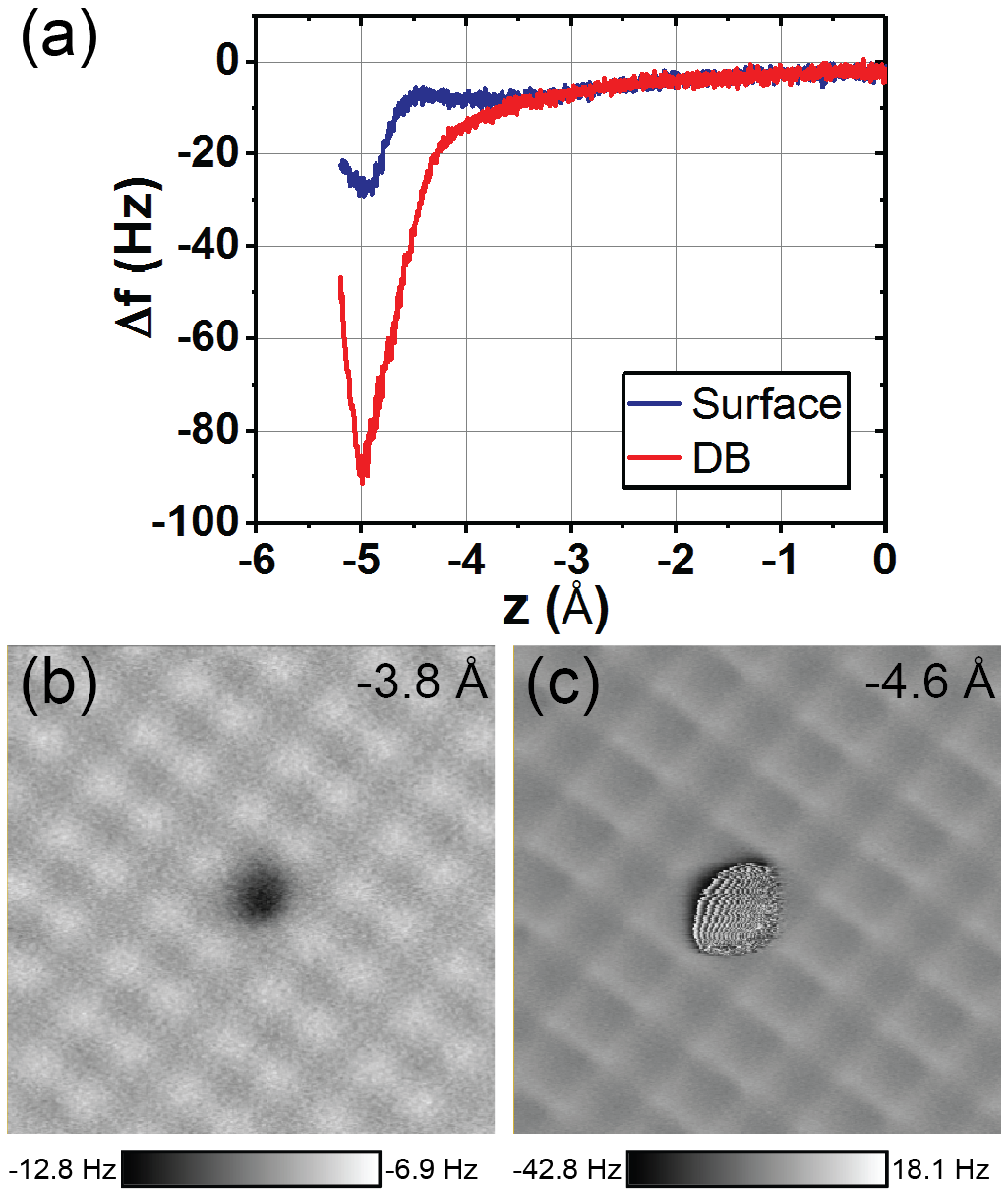}
	\caption{ \textbf{NC-AFM characterization of a single DB on the H-Si(100)-2x1 surface using a H-functionalized tip.} 
		(a) $\Delta$f(z) curves recorded on the H-Si surface (blue curve) and on the silicon DB (red curve). (3$\times$3)$nm^2$ frequency shift maps of a DB on the H-Si surface at relatively large (b) and small (c) tip-sample distances respectively. All data was acquired at 0~V with an oscillation amplitude of 1~\AA{}.}
	\label{Fig4}
\end{figure}

Figure \ref{Fig4}-a shows force curves acquired using a H-functionalized tip above a surface hydrogen atom (blue curve) and a single silicon DB (red curve). These force curves were recorded subsequently with Z=0~\AA{} corresponding to the tip position defined by the STM imaging set points (30~pA and +1.3~V) before switching off the feedback loop. Hence, superposing the 2 curves allows direct comparison of the interaction force between tip-surface and tip-DB. We notice that for relatively large tip-sample distances, the two curves are almost identical. Only for small tip elevations (around -3.5~\AA{} in this example) is a difference seen, with the DB showing a much larger increase in attractive interaction with the tip. This indicates that short range forces are the main contributor to the interaction force. \cite{Such.2015} This is also consistent with the DB being a reactive chemical center on the chemically inert H-Si surface where deposited molecules can selectively adsorb.\cite{Piva.2005,Godlewski.2013} Similar to what was reported previously for the case of gold atoms adsorbed on NaCl over Cu(111),\cite{Gross.2009-1,Bocquet.2011} the short range electrostatic force due to the localized negative charge on the DB \cite{Haider.2009,Taucer.2014,Labidi.2015} is most likely the main contributor to the large tip-sample interaction on the DB. 

Figures \ref{Fig4}-b and c show frequency shift maps acquired at different tip elevations using a H-functionalized tip. At relatively large tip-sample distances (Figure \ref{Fig4}-b), each hydrogen atom decorating a silicon atom clearly appears and follows the dimer structure of the 2$\times$1 reconstruction. The DB arising from the desorbed hydrogen atom on the surface appears as a dark atom-sized protrusion, introducing a much higher tip-sample interaction force localized on the DB. As we get closer to the surface (Supplementary materials, Figure S2), an evolution from atomic to bond contrast is seen on the H-Si surface as discussed in detail in reference.\cite{Labidi.2017} In Figure \ref{Fig4}-c, as a result of the larger attractive forces, the DB appears as a enlarged feature. The perturbations seen inside are an artifact in the excitation channel due to the inability of the feedback loop to maintain a constant oscillation amplitude (Supplementary materials, Figure S3).

We note here that high resolution bond contrast imaging is rendered possible thanks to the passivation of the tip apex with a hydrogen atom.\cite{Gross.2009,Labidi.2017} The later can be attracted to form a covalent bond with the silicon DB, but only at very small tip-sample elevations. This shows that the H-functionalized tip is robust and can be used to image reactive adsorbates or surface defects. \\

\noindent \textbf{Altering engineered on-surface DB-based quantum structures}
Recent works have shown that a DB can behave as a single atom quantum dot. Additionally, using atom-by-atom lithography with the STM tip, the coupling between DBs can be exploited to create functional DB structures such as QCA circuits, binary wires and logic gates.\cite{Haider.2009,Wolkow.2014,Kolmer.2015}
For large many-atom circuits this necessitates a precise control of desorption, which is difficult to achieve and has not been reported for more than a few DBs so far. Hence, a technique to correct or change multi-DB structures is highly desirable. Additionally, capping DBs would allow modulating the engineered quantum states from coupled DBs.\cite{Schofield.2013}

\begin{figure}
	\centering
	\includegraphics[]{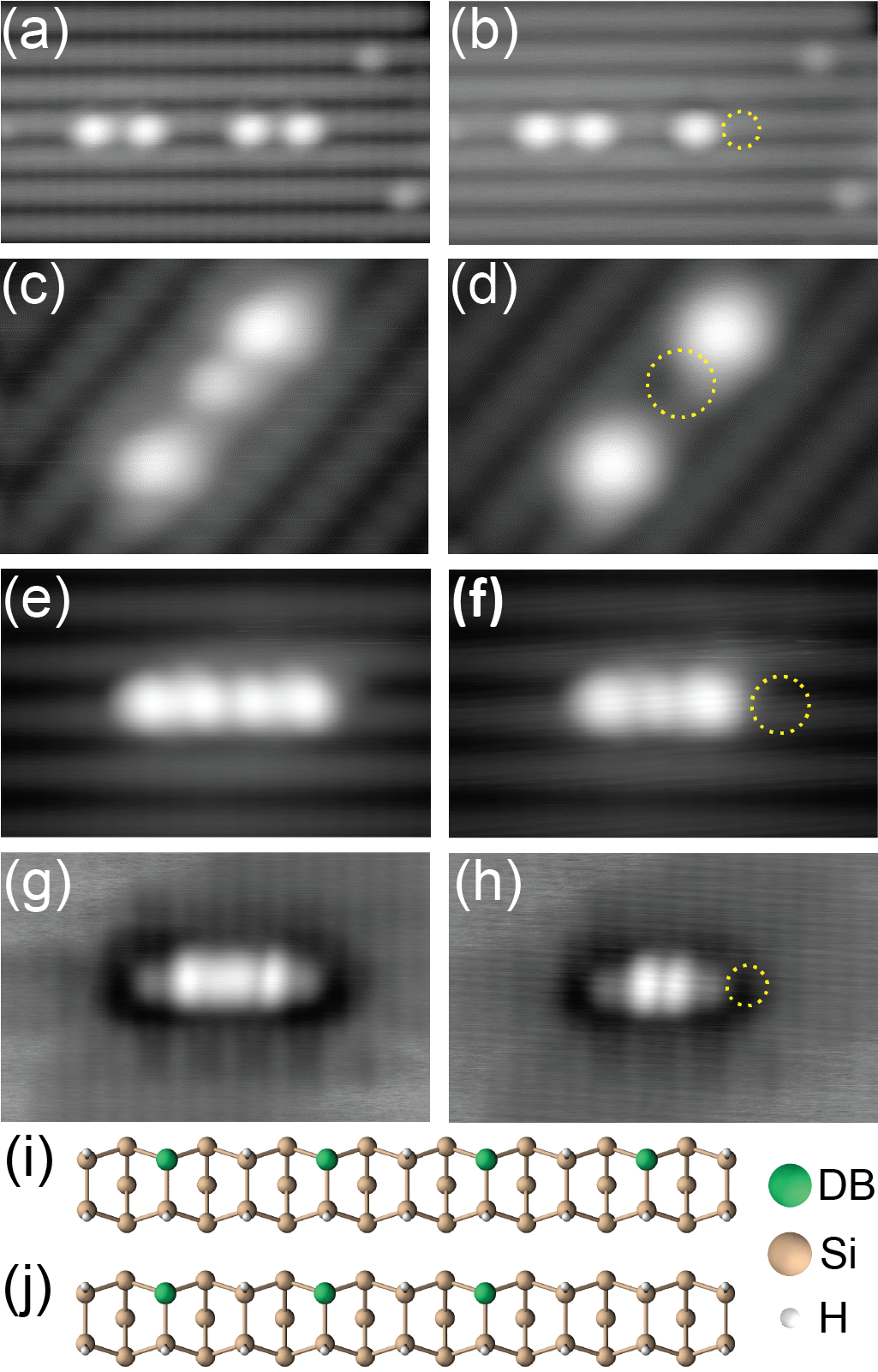}
	\caption{\textbf{Altering coupling and artificial molecular orbitals in multi-DB structures} (a) Two pairs of coupled DBs on the H-Si(100) surface arranged along a same dimer row. (b) Image of the same area after the mechanically induced capping of the far right DB in (a). 
		(c) A (3$\times$2)$nm^2$ STM image of three tunnel-coupled DBs. (b) The same area after erasing the middle DB in (c). Constant current images (a) to (d) were acquired at -1.8~V and 50~pA. (e)-(f) Filled (-2.0~V, 50~pA) and (g)-(h) empty (+1.4~V, 50~pA) states STM images of a DB wire, respectively, before and after erasing the far right DB in (e). 3d models of the four (i) and three (j) DB wire. Positions of erased DBs are indicated by dotted circles.}
	\label{Fig5}
\end{figure}

Figure \ref{Fig5}-a shows a filled state STM image of two separate pairs of coupled DBs along the same dimer row. We note here the enhanced STM contrast characteristic of a H-functionalized tip. In Figure \ref{Fig5}-b, the right side DB was selectively capped using the mechanically induced H-Si covalent bonding described in the previous section. We notice in Figure \ref{Fig5}-b the change in the STM contrast as previously shown in Figures \ref{Fig1} and \ref{Fig3}. Additionally, the now single DB on the right side of the image appears as a bright protrusion surrounded with a small dark halo, while the appearance of the two other coupled DBs shows no change.\cite{Haider.2009} 

A similar experiment is shown in Figure \ref{Fig5}-c where three tunnel coupled DBs are imaged using a H-functionalized tip. 
The central DB was then erased, the tip re-functionalized by picking up another hydrogen atom, and the remaining two DBs re-imaged in Figure \ref{Fig5}-d. Using an equivalent hydrogen tip for the before and after images highlights that changes in brightness are the result of coupling alterations, not simply a change of the terminating atom.  

Figure \ref{Fig5}-e shows a filled state image of four DBs along the same dimer row with each DB separated by a single H atom as illustrated in the 3D model of Figure \ref{Fig5}-f. 
In the corresponding empty states image (Figure \ref{Fig5}-g), a more complex structure is seen with four additional bright protrusions between the visible end atoms. Theoretical analysis from Schofield \textit{et al.}\cite{Schofield.2013} have reported similar results, explaining the extra protrusions as an excited state molecular orbital from wave function overlap of multi-DB systems. We show here an active modification of these artificial molecular orbitals through the controlled mechanical covalent bonding of a Si atom (DB) with a hydrogen atom on the tip apex causing nodes to disappear.  
Figure \ref{Fig5}-f and h respectively show the altered filled and empty state molecular orbitals from erasing the far right DB in Figure \ref{Fig5}-e. The filled state image shows up as three bright protrusions corresponding to three DBs, whereas the empty state image has been altered to now only have 2 bright protrusions instead of the prior 4. 
We note that in the example of Figure \ref{Fig5}-e to h, DB structures were imaged using a non-functionalized tip both before and after alteration. This further highlights that changes in the coupling between DBs visible from the different additional nodes that appear/disappear is not due to changes in the tip, but rather the result of erasing a DB with a hydrogen on the tip apex.  

Through the examples of Figure \ref{Fig5}, we can see how the controlled mechanically induced H and Si covalent bonding allows the non-destructive editing of a DB structure. This technique could be further applied to actuation of more complicated DB based patterns and elements as well, with erasure of a DB acting as a type of switch.  

\section{Conclusion and outlook}
To summarize, we showed that following a tip induced desorption, a hydrogen atom can be deposited on the surface or transfered to the tip apex resulting in a H-functionalized tip. The physisorbed single hydrogen atom on the chemically inert H-Si surface could be stably imaged in STM and AFM modes.
The H-functionalized tip was used to (i) characterize silicon dangling bonds and (ii) to mechanically induce the covalent bonding of single hydrogen and silicon atoms. We showed the potential of this mechanically induced reaction to precisely modify multiple DB structures such as coupled DBs and artificial molecular states.

\section{Methods}
Experiments were carried out using a commercial (Omicron) LT-STM/AFM system operated at 4.5~K.
We used qPlus sensors exhibiting a quality factor $Q\simeq 30,000$ and a resonance frequency $f_0\simeq 25 kHz$. Tips were direct current etched in a NaOH solution from a 50~$\mu$m thick polycrystalline wire. In ultra high vacuum (UHV), tips were first cleaned from their oxide layer by a series of electron beam heating treatments, followed by field evaporation in a field ion microscope (FIM). Then, they were further sharpened using a FIM nitrogen etching process to ensure small tip radius of curvature.\cite{Labidi.2015-1} The sensor was equipped with a separate wire for tunneling current to minimize cross-talk problems.\cite{ Majzik.2012} Additionally, all AFM data was recorded at 0~V to avoid both coupling between frequency shift and tunnel current, as well as imaging artifacts such as phantom force.\cite{Nony.2016,Weymouth.2011} 

In situ, tips were further processed to obtain artifact free images by first creating a bare silicon patch through tip induced hydrogen desorption, then by controlled nano-indentation on the created patch. This procedure usually results in a clean and stable tip.\cite{Labidi.2017} 

We used highly arsenic-doped ($\sim\, 1.5\times 10^{19}\, atom\, cm^{-3})$ silicon (100) samples. Following 12 hours degassing at $\sim 600\,^{\circ} C$ in UHV, samples underwent a series of resistive flash anneals at $1250\,^{\circ} C$.\cite{Labidi.2015} Samples were then exposed to atomic hydrogen while being kept at $\sim 320\,^{\circ} C$ to ensure a 2$\times$1 reconstruction.

To minimize drift during AFM image acquisition, the tip was left to settle for about 12 hours after approach to allow piezo scanner stabilization. Additionally, an atom tracking program implemented in the Nanonis control electronics was used. 

\section{Acknowledgment}
We thank M. Cloutier for his technical expertise.
We thank NRC, NSERC, QSi, and AITF for financial support.

\pagebreak
\widetext
\begin{center}
	\textbf{Supplementary Materials:\\ Atomic White-Out: Enabling Atomic Circuitry Through Mechanically Induced Bonding of Single Hydrogen Atoms to a Silicon Surfacet}
\end{center}
\setcounter{equation}{0}
\setcounter{figure}{0}
\setcounter{table}{0}
\setcounter{page}{1}
\makeatletter
\renewcommand{\theequation}{S\arabic{equation}}
\renewcommand{\thefigure}{S\arabic{figure}}
\renewcommand{\bibnumfmt}[1]{[S#1]}
\renewcommand{\citenumfont}[1]{S#1}

\begin{figure}[htbp]
	\centering
	\includegraphics[width=1.0 \textwidth]{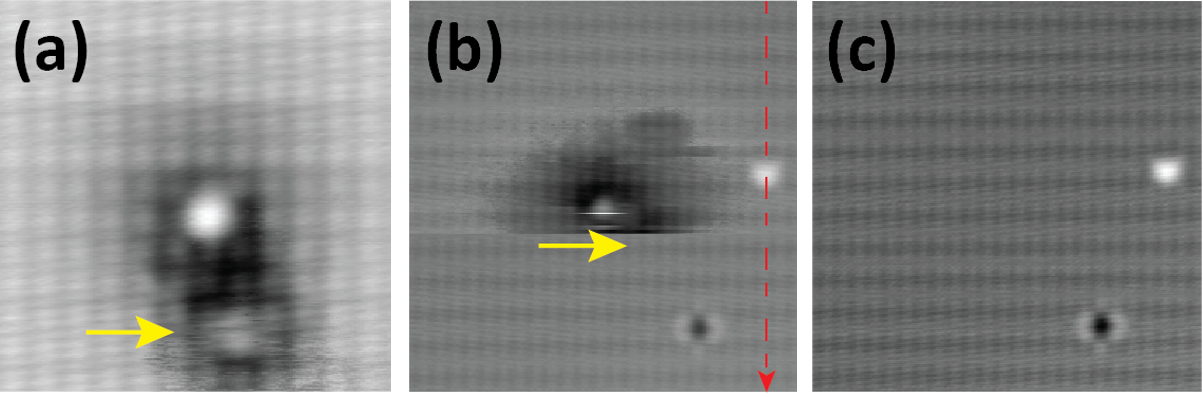}
	\caption{(a) Single hydrogen atoms physisorbed on the chemically inert H-Si(100) surface could be stably imaged in filled states at low voltage (+1.3~V). However, when the scanning voltage is increased to +1.7V in (b), the hydrogen atom is dragged by the tip which resulted in the capping of the DB during the STM image as indicated by a change in contrast midway through the image and confirmed by a subsequent STM image of the same area (c). (b) and (c) are larger area (10$\times$10)$nm^2$) images of the area in (a). The location of the atomic hydrogen is marked with an arrow.}
	\label{FigS1}
\end{figure}

\begin{figure}[htbp]
	\centering
	\includegraphics[width=1.0\textwidth]{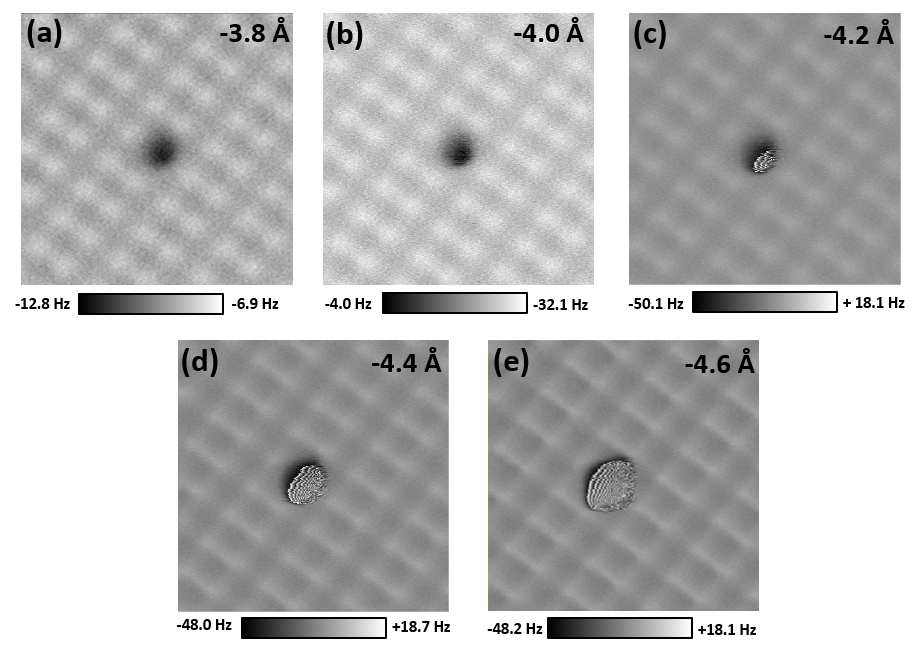}
	\caption{(a)-(e) Series of raw (3$\times$3)$nm^2$ NC-AFM frequency shift maps of H-Si(100) surface at different tip-sample elevations. Images were recorded at 0~V and with an oscillation amplitude of 1~\AA{}. We see the evolution from atomic to chemical bond contrast on the H-Si surface. For smaller tip elevations, much higher interaction force is seen on the DB than elsewhere on the surface. Z=0~\AA{} corresponds to the tip position defined by the STM imaging set points (30~pA and +1.3~V) before switching off the feedback loop.}
	\label{FigS2}
\end{figure}

\begin{figure}[htbp]
	\centering
	\includegraphics[width=1.0 \textwidth]{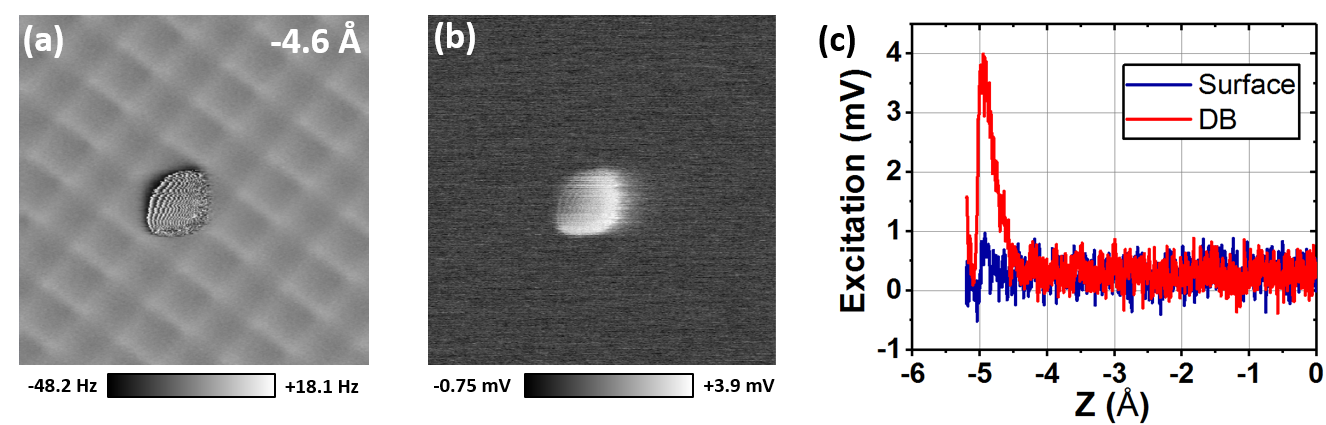}
	\caption{(a) NC-AFM frequency shift map of a single DB at small tip-sample distance (-4.6 \AA{}) and (b) corresponding simultaneously obtained excitation channel map. (c) Superposed excitation versus tip elevation curves recorded on the same DB (red curve) and on the H-Si surface (blue curve) }
	\label{FigS3}
\end{figure}

\end{document}